\documentclass[12pt,a4paper,dvips,rotating]{article}
\usepackage{a4p}

\usepackage{times,psmath}
\usepackage{cite,mcite}
\usepackage{graphicx}
\usepackage{physics }
\usepackage{ams_title,ifthen,Lep}
\journalname{Phys. Lett. B}
\date{}
\myjournaln
\graphicspath{{.}}
\newlength{\capindent}
\newlength{\capwidth}
\newlength{\figwidth}
\newlength{\figheight}
\newcommand{\icaption}[2][!*!,!]{\hspace*{\capindent}%
  \begin{minipage}{\capwidth}
    \ifthenelse{\equal{#1}{!*!,!}}%
      {\caption{#2}}%
      {\caption[#1]{#2}}
  \end{minipage}}
\newcommand{\pho}{\phantom{0}}
\newcommand{\dpho}{\phantom{.0}}

\newcommand{\TM}{\ensuremath{{\Theta_\mathrm{M}}}}

\newcommand{\R}{\ensuremath{{\mathrm{R}}}}
\begin{document}
\begin{titlepage}
\title{Protons in Near Earth Orbit}
\author{The AMS Collaboration}
%
%
\begin{abstract}
The proton spectrum in the kinetic energy range 0.1 to
200\,\GeV\ was measured by 
the Alpha Magnetic Spectrometer (AMS)
during space shuttle flight STS--91 at an
altitude of 380\,km.  Above the geomagnetic cutoff the observed spectrum is
parameterized by a power law.  Below the geomagnetic cutoff a substantial
second spectrum was observed concentrated at equatorial latitudes with a flux 
$\sim$70\,m$^{-2}$sec$^{-1}$sr$^{-1}$.
Most of these second spectrum protons
follow a complicated trajectory and
originate from a restricted geographic region.
\end{abstract}
%
%
\submitted
\end{titlepage}
%
%

\section*{Introduction}
Protons are the most abundant charged particles in space. The study
of cosmic ray protons improves the understanding of the interstellar
propagation and acceleration of cosmic rays.

There are three distinct regions in space where protons have been
studied by different means:
\begin{itemize}
\item 
The altitudes of 30--40\,km above the Earth's surface. This region has been
studied with balloons for several decades. Balloon experiments have made
important contributions to the understanding of the primary cosmic ray
spectrum of protrons and the behavior of atmospheric secondary particles in
the upper layer of the atmosphere.
\item
The inner and outer radiation belts, which extend from altitudes of about
1000\,km up to the boundary of the magnetosphere. Small size detectors on
satellites have been sufficient to study the high intensities in the
radiation belts.
\item
A region intermediate between the top of the atmosphere and the inner
radiation belt.  The radiation levels are normally not very high, so
satellite-based detectors used so far, \ie\/ before AMS, have not been
sensitive enough to systematically study the proton spectrum in
this region over a broad energy range.
\end{itemize}
Reference~\mcite{REF1}
includes some of the previous studies.  The primary
feature in the proton spectrum observed near Earth 
is a low energy drop off in the flux, 
known as the geomagnetic cutoff.  This cutoff occurs at kinetic
energies ranging from $\sim$10\,\MeV\ to $\sim$10\,\GeV\ 
depending on the latitude and
longitude.  Above cutoff, from $\sim$10 to $\sim$100\,\GeV, 
numerous measurements
indicate the spectrum falls off according to a power law.

The Alpha Magnetic Spectrometer (AMS)~\cite{oldnim} is a high energy physics
experiment scheduled for installation on the International Space Station.
In preparation for this long duration mission, AMS flew a precursor mission
on board the space shuttle Discovery during flight STS--91 in June 1998.  In
this report we use the data collected during the flight to study the cosmic
ray proton spectrum from kinetic energies of 0.1 to 200\,\GeV, taking
advantage of the large acceptance, the accurate momentum resolution, the
precise trajectory reconstruction and the good particle identification
capabilities of AMS.

The high statistics ($\sim10^7$) available allow the variation of the spectrum
with position to be measured both above and below the geomagnetic cutoff.
Because the incident particle direction and momentum were accurately
measured in AMS, it is possible to investigate the origin of protons below
cutoff by tracking them in the Earth's magnetic field.

\section*{The AMS Detector}
The major elements of AMS  
as flown on STS--91
consisted of a permanent magnet, a tracker, time of flight
hodoscopes, a Cerenkov counter and anticoincidence 
counters~\mcite{REF3}. 
The permanent magnet had the shape of a cylindrical shell 
with inner diameter 1.1\,m, length 0.8\,m
and
provided a central dipole field of 0.14\,Tesla across the magnet bore
and an analysing power, $BL^2$, of 0.14\,Tm$^2$ parallel to the
magnet, or z--, axis.  
The six layers of double sided silicon tracker were arrayed
transverse to the magnet axis. 
The outer layers  were just outside the magnet cylinder.
The tracker measured the
trajectory of relativistic singly charged particles with an accuracy 
of 20\,microns
in the bending coordinate and 33\,microns 
in the non-bending coordinate, as well as
providing multiple measurements of the energy loss.  
The time of flight system had two planes at each end of the magnet,
covering the outer tracker layers.
Together the four planes 
measured singly charged particle transit times with
an accuracy of 120\,psec and also yielded multiple energy loss measurements.
The Aerogel Cerenkov counter ($n=1.035$) was used to
make independent velocity measurements to separate low energy protons from
pions and electrons.  A layer of anticoincidence scintillation counters
lined the inner surface of the magnet.  Low energy particles were
absorbed by thin carbon fiber shields.
In flight the AMS positive z--axis pointed out of the shuttle payload bay.

For this study the
acceptance was restricted to events with an incident angle within $32^\circ$
of the positive z--axis of AMS 
and data from two periods are
included.  In the first period the z--axis was pointing within 
$1^\circ$ of the zenith. 
Events from this period are referred to as ``downward'' going.  In
the second period the z--axis pointing was within $1^\circ$ of the nadir.  
Data from this period are referred to as ``upward'' going. 
The orbital inclination was $51.7^\circ$ and the geodetic
altitude during these two periods ranged from 350 to 390\,km.
Data taken while
orbiting in or near the South Atlantic Anomaly were excluded.

The response of the detector was simulated using the AMS detector
simulation program, based on the GEANT package~\cite{REF4}. 
The effects of energy
loss, multiple scattering, interactions, decays and the measured
detector efficiency and resolution were included.

The AMS detector was extensively calibrated at two accelerators:
at GSI, Darmstadt, with helium and carbon beams at 600 incident angles and
locations and $10^7$ events, and at the CERN proton-synchrotron in the energy
region of 2 to 14\,\GeV, with 1200 incident angles and locations 
and $10^8$ events.  This ensured that the performance of the 
detector and the analysis procedure were thoroughly understood.

\section*{Analysis}
Reconstruction of the incident particle type, energy and direction
started with a track finding procedure which included cluster finding,
cluster coordinate transformation and pattern recognition. The track was
then fit using two independent algorithms~\cite{GEANE,trfit1}. 
For a track to be
accepted the fit was required to include at least 4 hits in the bending
plane and at least 3 hits in the non-bending plane. 

The track was then extrapolated to each time of flight plane and
matched with the nearest hit if it was within 60\,mm.  Matched hits were
required in at least three of the four time of flight planes.  The
velocity, $\beta=\mathrm{v}/c$, 
was then obtained using this time of flight  
information and the trajectory. For events which passed through the
Cerenkov counter sensitive volume an independent velocity measurement, 
$\beta_C$,
was also determined.  To obtain the magnitude of the particle charge, $|Z|$, 
a set of reference distributions of energy losses in both the time of flight
and the tracker layers were derived from calibration measurements made at
the CERN test beam interpolated via the Monte Carlo method.  For
each event these references were fit to the measured energy losses using a
maximum likelihood method.  The track parameters were then refit with the
measured $\beta$ and $Z$ 
and the particle type determined from the resultant 
$Z$, $\beta$, $\beta_C$ and rigidity,  $\R=pc/|Z|e$\,(GV).

As protons and helium nuclei are the dominant components in cosmic
rays, after selecting events with $Z = +1$ the proton sample has only minor
backgrounds which consist of charged pions and deuterons.  
The estimated fraction of
charged pions, which are produced in the top part of AMS, with energy
below 0.5\,\GeV\ is 1\,\%.  Above this energy the fraction decreases rapidly
with increasing energy.  The deuteron abundance in cosmic rays above the
geomagnetic cutoff is about 2\,\%.  To remove low energy charged pions 
and deuterons the measured mass was required to
be within 3 standard deviations of the proton mass.  This rejected 
about 3\,\%
of the events while reducing the background contamination to negligible
levels over all energies.

To determine the differential proton fluxes from the measured
counting rates requires the acceptance to be known as a function of the
proton momentum and direction.  Protons with different momenta and
directions were generated via the Monte Carlo method, 
passed through the AMS detector
simulation program and accepted if the trigger and
reconstruction  requirements were satisfied as for the data.  The
acceptance was found to be 0.15\,m$^2$sr on average, varying from 
0.3 to 0.03\,m$^2$sr with 
incident angle and location and only weakly momentum dependent. These
acceptances were then corrected following an analysis of unbiased trigger
events.  The corrections to the central value are shown in 
Table~\ref{sys1} together
with their contribution to the total systematic error of 5\,\%.
\begin{table}[h]
\begin{center}
\begin{tabular}{|ll|c|c|}
\hline
\multicolumn{2}{|l|}{Correction}&Amount&Uncertainty\\
\hline
\multicolumn{2}{|l|}{Trigger:}&              & \\
\hspace{5mm}
 &4--Fold Coincidence&             \llap{--\,}3& 1.5\\
 &Time of Flight Pattern&          \llap{--\,}4& 2\dpho\\
 &Tracker Hits&                    \llap{--\,}2& 1\dpho\\
 &Anticoincidence&                            0& 1\dpho\\
\hline
\multicolumn{2}{|l|}{Analysis:}&             &          \\
 &Track and Velocity Fit&          \llap{--\,}2& 1.5\\
 &Particle Interactions&           \llap{$+$\,}1& 1.5\\
 &Proton Selection&                \llap{--\,}2& 2\dpho\\
\hline
\multicolumn{2}{|l|}{Monte Carlo Statistics}& 0& 2\dpho\\
\hline
\multicolumn{2}{|l|}{Differential Acceptance Binning}& 0& 2\dpho\\
\hline
\multicolumn{2}{|l|}{\bf Total}&\bf \llap{--\,1}2&\bf 5\dpho\\
\hline
\end{tabular}
\end{center}
\caption{\label{sys1}Acceptance corrections and their systematic uncertainties,
in percent}
\end{table}

\begin{figure}[ht]
  \begin{center}    
    \includegraphics[width=\figwidth]{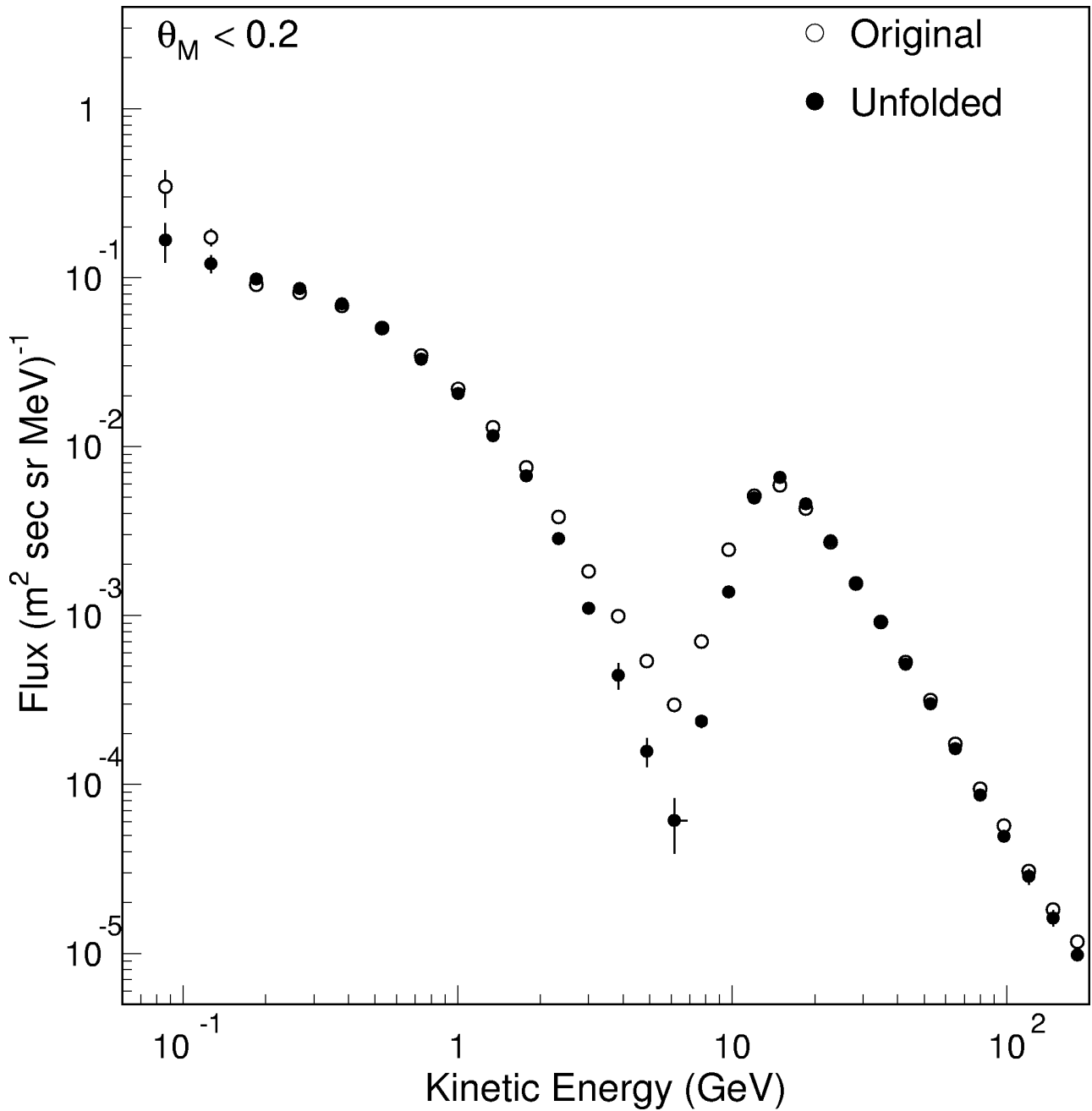}
  \end{center}
  \caption{The proton differential flux in the equatorial region. 
 Open circles show the measured distribution, 
 filled circles are the data after unfolding.
  \label{unfolding}}
\end{figure}

To obtain the incident differential spectrum from the measured
spectrum, the effect of the detector resolution was unfolded using
resolution functions obtained from the simulation. These functions were
checked at several energy points by test beam measurements.  The data were
unfolded using a method based on Bayes' 
theorem~\cite{REF7,blobel85}, 
which used an iterative
procedure (and not a ``regularized unfolding'') to overcome instability of
the matrix inversion due to negative terms. 
Fig.~\ref{unfolding} compares the differential
proton spectrum before and after unfolding in the geomagnetic equatorial
region, defined below.

\section*{Results and Interpretation}

The differential spectra in terms of kinetic energy for downward
and upward going protons integrated over incident angles within 32$^\circ$ 
of the AMS z--axis, which was within 1$^\circ$ of the zenith or nadir, 
are presented in Fig.~\ref{spectra}
and Tables~\ref{fluxdownlo}--\ref{fluxup}.
The results have been separated
according to the absolute value of the corrected geomagnetic 
latitude~\cite{CGC},
$\TM$ (radians), at which they were observed.  
Figs.~\ref{spectra}a, b and c clearly show the effect
of the geomagnetic cutoff and the decrease in this cutoff with increasing
$\TM$.  The spectra above and below cutoff differ.  The spectrum
above cutoff is refered to as the ``primary'' spectrum 
and below cutoff as the ``second'' spectrum.

\subsection*{I. Properties of the Primary Spectrum}
The primary proton spectrum may be parameterized
by a power law in
rigidity, $\Phi_0 \times \R^{-\gamma}$.  
Fitting~\cite{blobel85} the measured spectrum over the rigidity
range $10 < \R < 200$\,GV, \ie\/ well above cutoff, yields:
$$
\gamma = 2.79 \pm 0.012\,(\mbox{fit}) \pm 0.019\,(\mbox{sys}),
$$
$$
\Phi_0 = 
16.9 \pm 0.2\,(\mbox{fit}) 
     \pm 1.3\,(\mbox{sys}) 
     \pm 1.5\,(\gamma)\, 
\frac{\mathrm{GV}^{2.79}}{\mathrm{m^2sec\,sr\,MV}}\,.
$$
The systematic uncertainty in  $\gamma$ 
was estimated from the uncertainty
in the acceptance (0.006), the dependence of the resolution function on the
particle direction and track length within one sigma (0.015),  variation of
the tracker bending coordinate resolution by $\pm 4$\,microns (0.005) 
and variation of
the selection criteria (0.010).  The third uncertainty quoted for $\Phi_0$
reflects the systematic uncertainty in $\gamma$.

\subsection*{II. Properties of the Second Spectrum}
\begin{figure}[ht]
  \begin{center}               
    \includegraphics[height=\figheight]{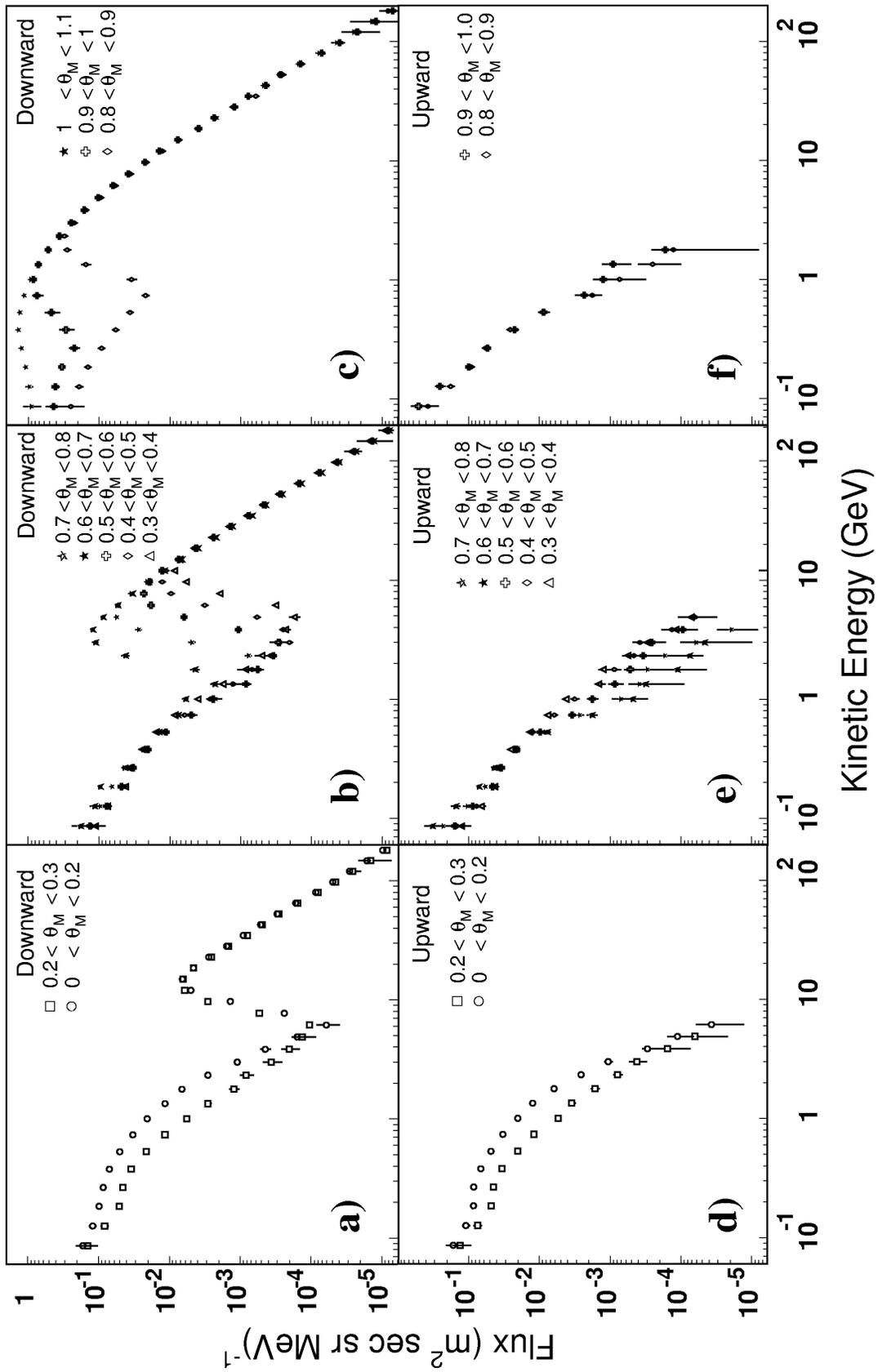}
  \end{center}
  \caption{Flux spectra for 
           a,b,c) downward and d,e,f) upward going protons
           seperated according to the geomagnetic latitude, $\TM$,
           at which they were detected.
  \label{spectra}}
\end{figure}

\begin{figure}[ht]
  \begin{center}                       
    \includegraphics[height=\figheight]{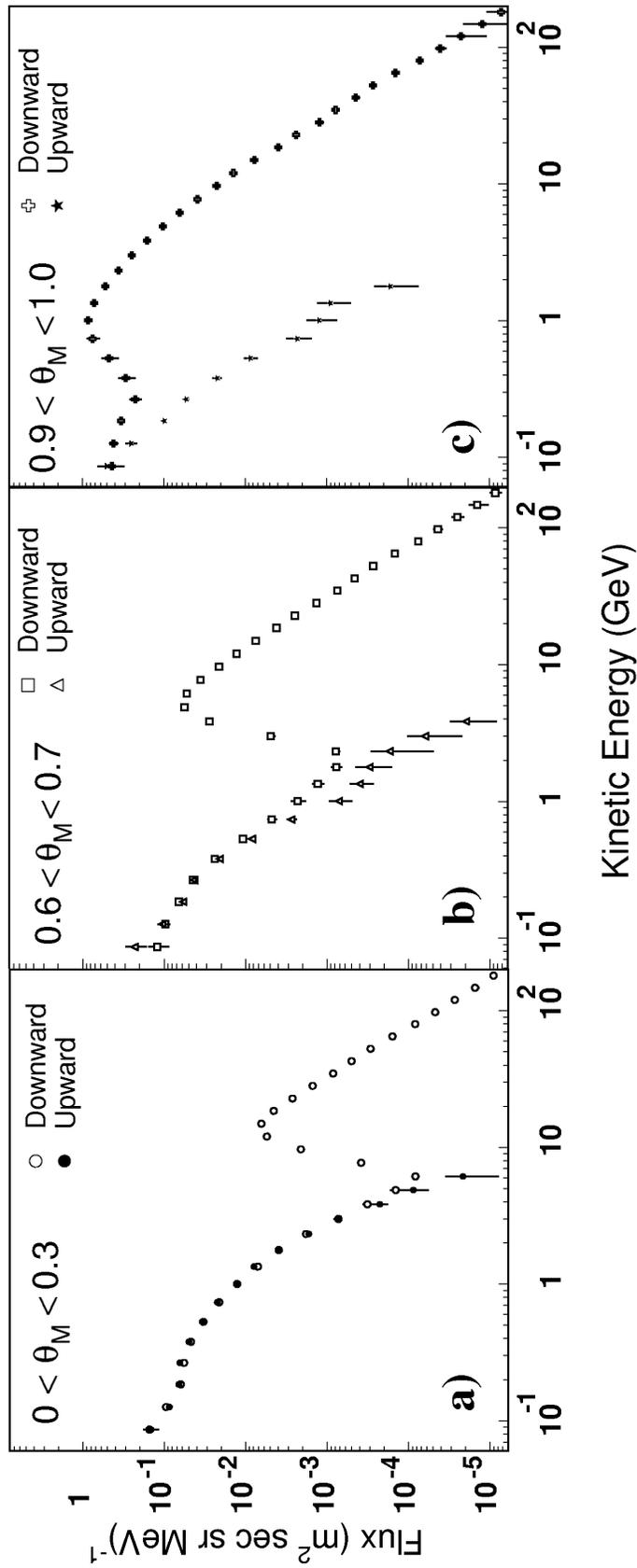}
  \end{center}
  \caption{Comparison of upward and downward second spectrum proton
           at different geomagnetic latitudes.  As seen,
           below cutoff, the upward and downward fluxes
           agree in the range $0\le\TM<0.8$
           (see also Figs.~\ref{spectra}b, e).
  \label{updown}}
\end{figure}

As shown in Figs.~\ref{spectra}a, b, c, 
a substantial second spectrum of
downward going protons is observed for all but the highest geomagnetic
latitudes. 
Figs.~\ref{spectra}d, e, f show that a substantial second
spectrum of upward going protons is also observed
for all geomagnetic latitudes.
The upward and downward going protons of the second spectrum
have the following unique properties:
\begin{itemize}
\item[(i)] At geomagnetic equatorial latitudes, $\TM < 0.2$,
this spectrum extends from the lowest measured energy, 
0.1\,\GeV, to $\sim$6\,\GeV\
with a flux $\sim$70\,m$^{-2}$sec$^{-1}$sr$^{-1}$.
\item[(ii)]
As seen in Figs.~\ref{spectra}a, d, the second spectrum 
has a distinct structure near the geomagnetic equator: 
a change in geomagnetic latitude from 0 to 0.3 
causes the proton flux to drop by a factor of 2 to 3 depending on the energy. 
\item[(iii)]
Over the much wider interval $0.3<\TM<0.8$, the flux is nearly constant.
\item[(iv)]
In the range  $0\le\TM<0.8$,
detailed comparison in different 
latitude bands (Fig.~\ref{updown}) indicates 
that the upward and downward fluxes
are nearly identical, agreeing within 1\,\%.
\item[(v)]
At polar latitudes, $\TM  > 1.0$, 
the downward second spectrum (Fig.~\ref{spectra}c) 
is gradually obscured by the primary spectrum,
whereas the second spectrum of upward going protons  (Fig.~\ref{spectra}f)
is clearly observed.
\end{itemize}

\begin{figure}[ht]
  \begin{center}                    
    \includegraphics[width=\figwidth]{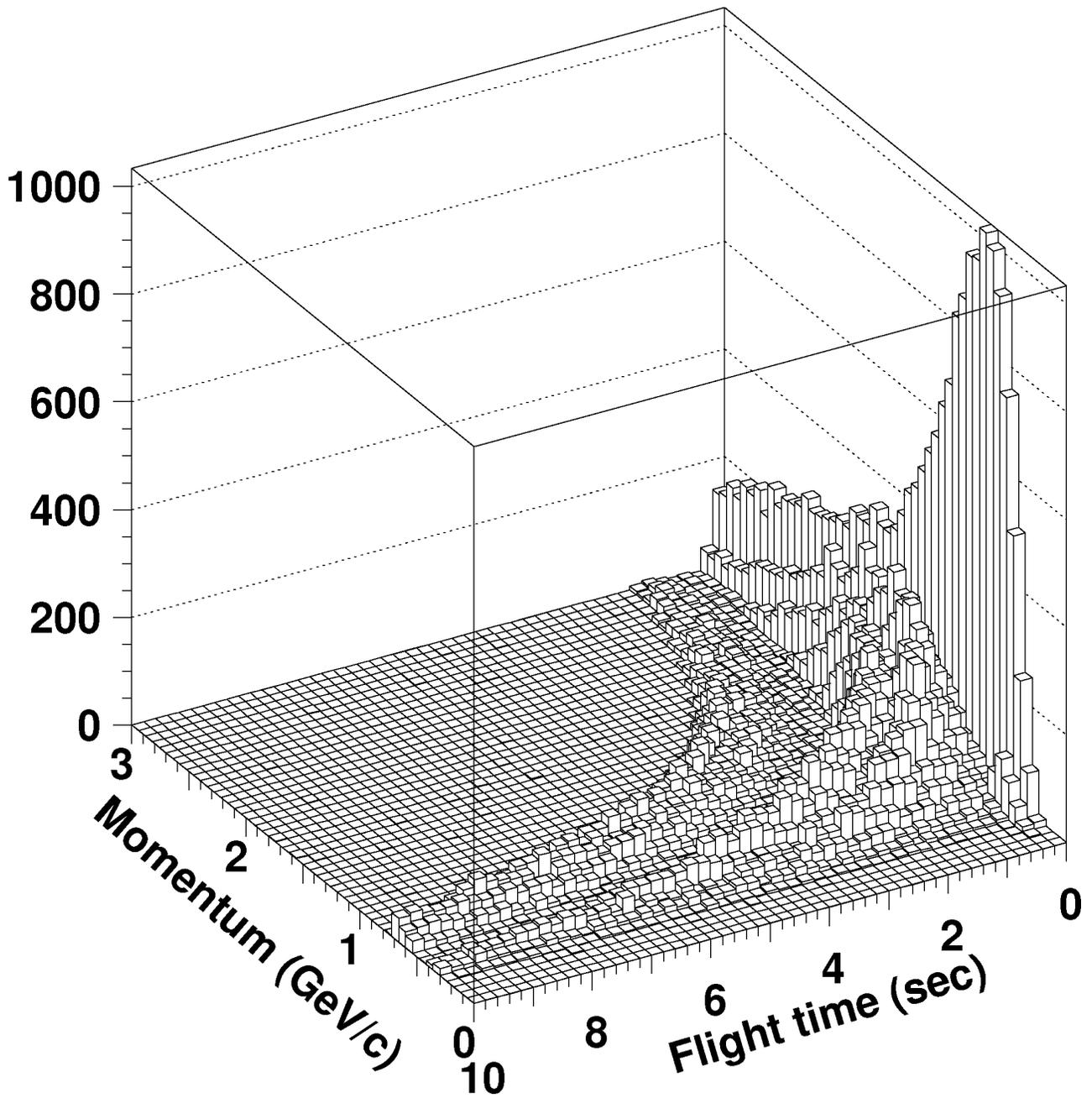}
  \end{center}
  \caption{The interval between production and detection, or flight time,
           versus momentum from the back tracing of protons detected
           in the region $\TM<0.3$.
  \label{pvslife}}
\end{figure}

To understand the origin of the second spectrum, 
we traced~\cite{REFA} 
back
$10^5$ protons 
from their measured incident angle, location and momentum,
through the geomagnetic 
field~\cite{REFB} 
for 10\,sec flight time
or until they impinged on the top of the 
atmosphere at an altitude of 40\,km, 
which was taken to be the point of origin. 
All second spectrum protons were found to originate in the atmosphere,
except for few percent of the total detected near the 
South Atlantic Anomaly (SAA). 
These 
had closed trajectories and hence 
may have been circulating for a very long time
and it is
obviously difficult to trace back to thier origin.
This type of trajectory was only observed
near the SAA, 
clearly influenced by the inner radiation belt.
To avoid confusion data taken in the SAA region were
excluded though the rest of the protons detected near the SAA
had characteristics as the rest of the sample.
Defining the flight time as the interval between production
and detection, Fig.~\ref{pvslife} 
shows the distribution of momentum versus flight time of the
remaining protons.

\begin{figure}[ht]
  \begin{center}                    
    \includegraphics[width=\figwidth]{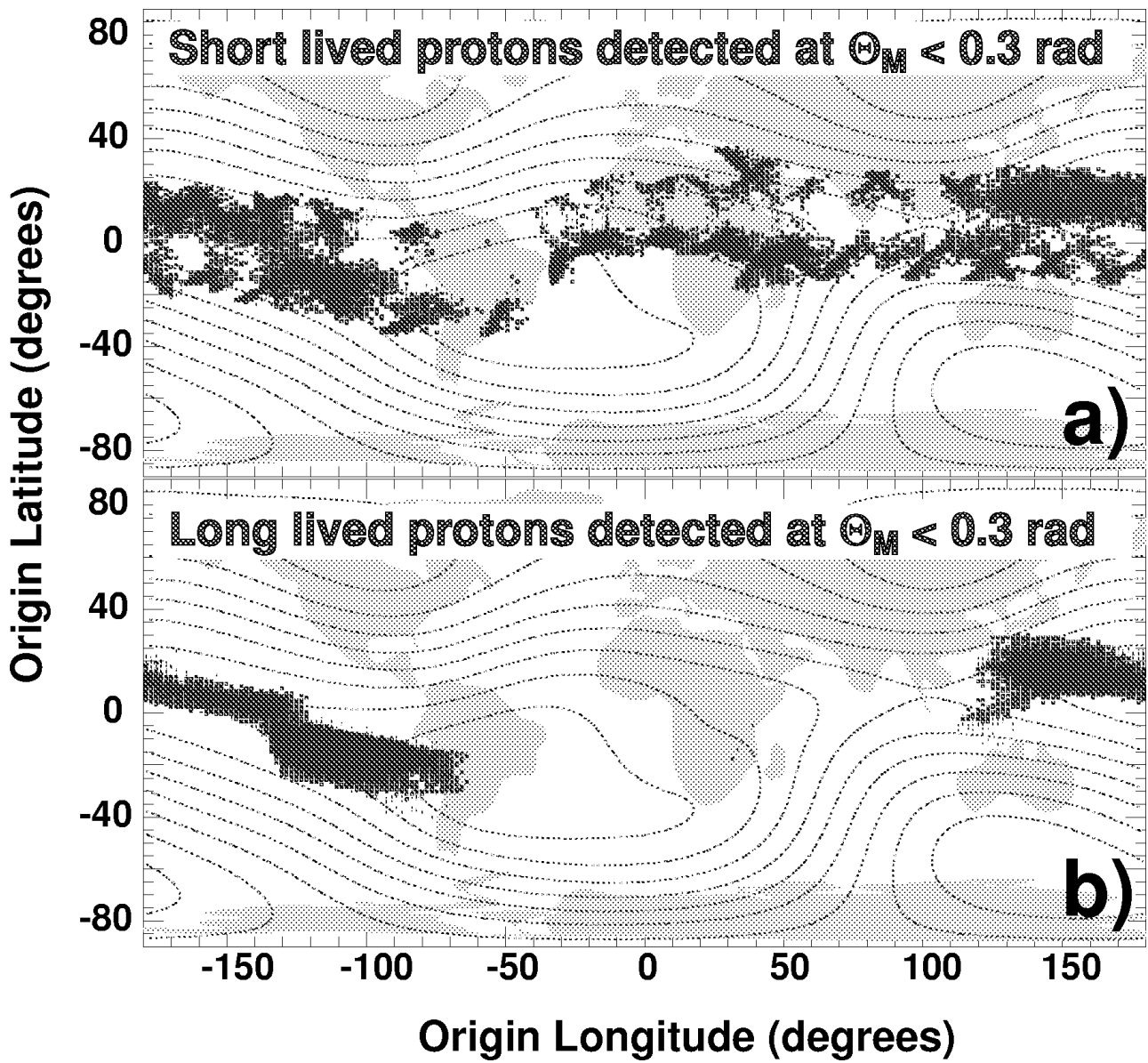}
  \end{center}
  \caption{The geographical origin of a) short--lived and b) long--lived
protons with $p<3$\,\GeV/c. The dashed lines indicate the geomagnetic
field countours at 380\,km.}
  \label{geog_origin}
\end{figure}

\begin{figure}[ht]
  \begin{center}                    
    \includegraphics[width=\figwidth]{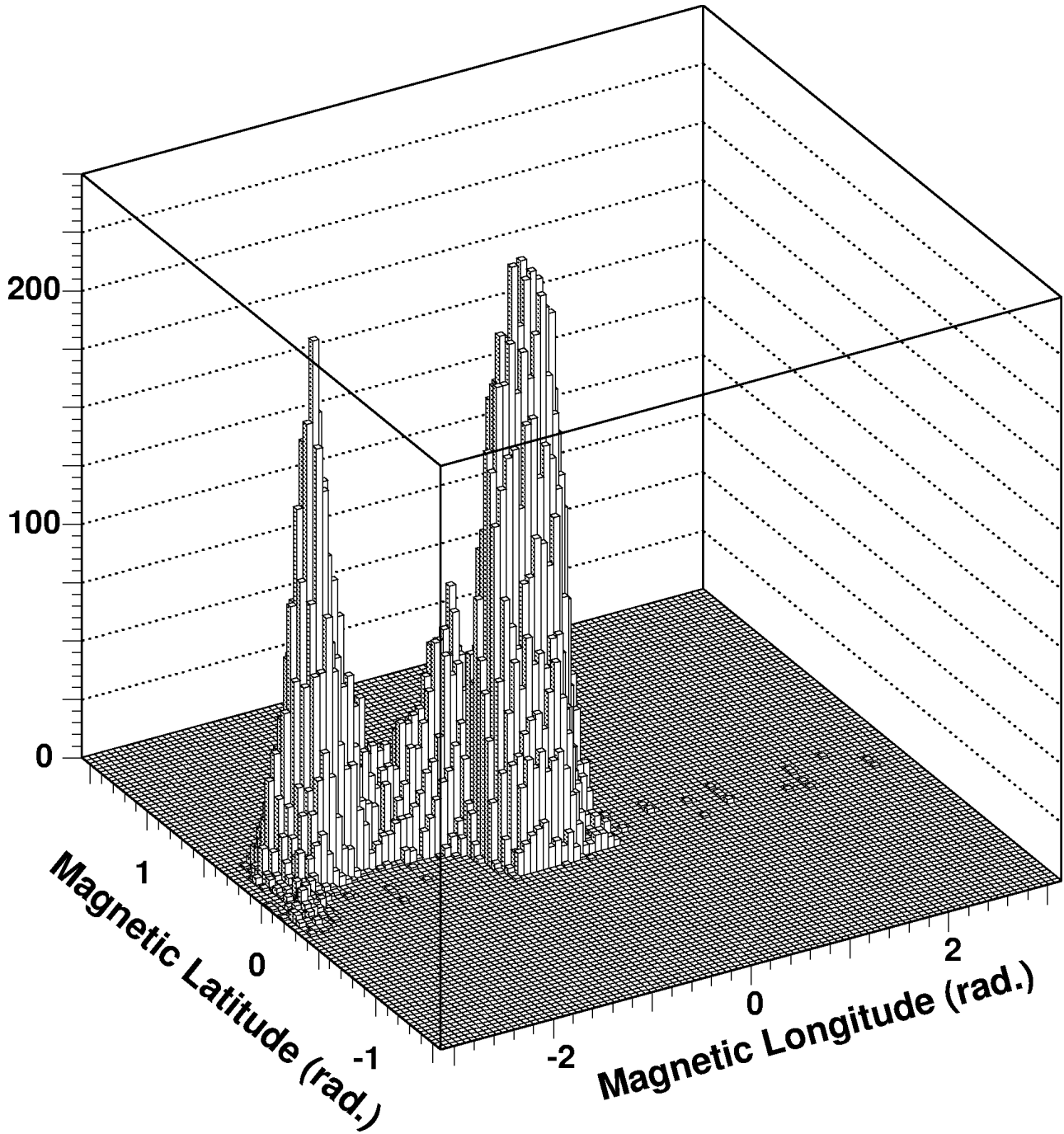}
  \end{center}
  \caption{The point of origin of long--lived protons 
           ($\TM<0.3, p<3$\,\GeV/c) in 
           geomagnetic coordinates.
  \label{geom_origin}}
\end{figure}

\begin{figure}[ht]
  \begin{center}                    
    \includegraphics[width=\figwidth]{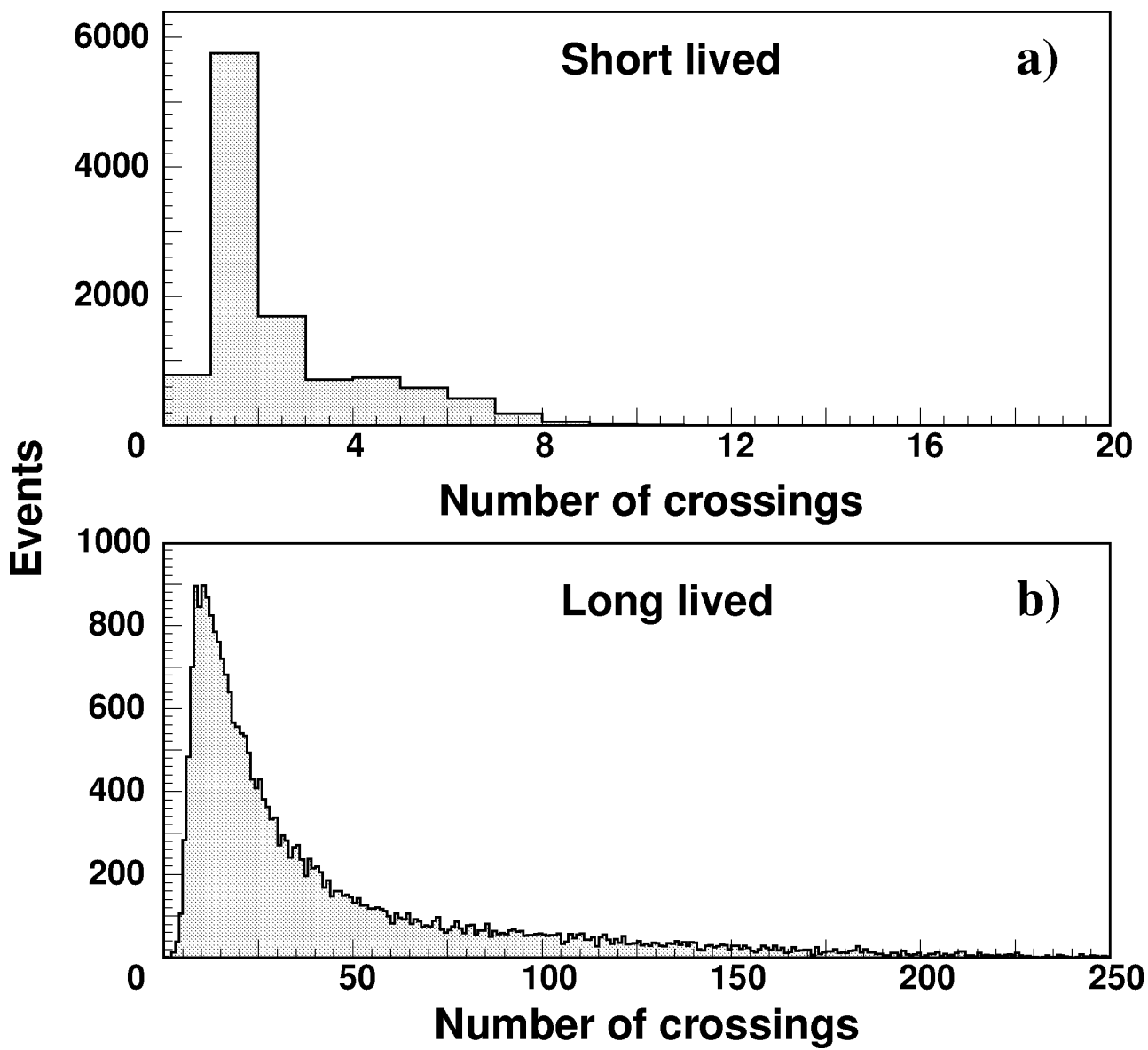}
  \end{center}
  \caption{Number of times the back traced trajectory
           crosses the geomagnetic equator for a) short--lived
           and b) long--lived protons ($\TM<0.3, p<3$\,\GeV/c).
  \label{equ_cross}}
\end{figure}

As seen in Fig.~\ref{pvslife},
the trajectory tracing shows that about 30\,\% of the detected protons
flew for less than 0.3\,sec before detection.  
The origin of these
``short--lived'' protons is distributed uniformly around the globe,
see Fig.~\ref{geog_origin}a, 
the apparent structure reflecting the orbits of the space shuttle.
In contrast, Fig.~\ref{geog_origin}b shows that the remaining 
70\,\% of protons with flight
times greater than 0.3\,sec, classified as ``long--lived'',
originate from a geographically restricted zone.
Fig.~\ref{geom_origin} shows the strongly peaked distribution
of the point of origin of these long--lived protons
in geomagnetic coordinates.
Though data is presented only for protons detected at $\TM<0.3$,
these general features hold true up to $\TM\sim0.7$.
Fig.~\ref{equ_cross} shows
the distribution of the number of geomagnetic equator crossings for
long--lived and short--lived protons.  
About 15\,\% of all the
second spectrum protons were detected on their first bounce over the
geomagnetic equator.

The measurements by AMS in near Earth orbit (at 380\,km
from the Earth's surface),
between the atmosphere and the radiation belt,
show that the particles in this region follow a
complicated path in the Earth's magnetic field.  
This behavior is different
from that extrapolated from satellite observations in the radiation belts,
where the protons bounce across the equator
for a much longer time.
It is also different
from that extrapolated  
from balloon observations in the upper layer of the atmosphere,
where the protons typically cross the equator once.
A striking feature of the second spectrum is that most
of the protons originate from a restricted geographic region.

%
%

\section*{Acknowledgements}
The support of 
INFN, Italy, 
ETH--Z\"urich, 
the University of Geneva,
the Chinese Academy of Sciences,
Academia Sinica and National Central University, Taiwan,
the RWTH--Aachen, Germany,
the University of Turku, the University of Technology of Helsinki, Finland,
the U.S.~DOE and M.I.T.,
CIEMAT, Spain, LIP, Portugal and IN2P3, France,
is gratefully acknowledged.

We thank Professors 
S.~Ahlen, 
C.~Canizares,
A.~De~Rujula,
J.~Ellis, 
A.~Guth,
M.~Jacob,
L.~Maiani,
R.~Mewaldt,
R.~Orava,
J.~F.~Ormes
and
M.~Salamon 
for helping us to initiate this experiment.

The success of the first AMS mission is due to
many individuals and organizations
outside of the collaboration.
The support of NASA
was vital in the inception, development and operation of the experiment.
The dedication of 
Douglas P.~Blanchard, 
Mark J.~Sistilli, 
James R.~Bates,
Kenneth Bollweg
and the NASA and Lockheed--Martin Mission Management team,
the support of the Max--Plank Institute for Extraterrestrial Physics,
the support of the space agencies from 
Germany (DLR),
Italy (ASI),
France (CNES) and
China 
and the support of CSIST, Taiwan, 
made it possible to complete this experiment on time.

The support of CERN and GSI--Darmstadt,
particularly of Professor Hans Specht and Dr.~Reinhard Simon
made it possible for us to calibrate the detector after the shuttle
returned from orbit.

The back tracing was made possible by the work of 
Professors E. Fl\"uckiger, D.~F.~Smart and M.~A. Shea.

We are most grateful to the STS--91 astronauts,
particularly to Dr.~Franklin Chang--Diaz 
who provided vital help to AMS during the flight.

%
\newpage
%
\bibliographystyle{amsstylem}
\bibliography{proton}
%
%
\newpage
\typeout{   }     
\typeout{Using author list for AMS paper 02}
\typeout{$Modified: Nov 99 by M. Capell $}
\typeout{!!!!  This should only be used with document option a4p!!!!}
%
%
%
%
%
%

\newcounter{tutetotcount}
\newcounter{tutetutecount}
\newcounter{tutecount}
\newcounter{namecount}
\newcommand{\tutenum}[1]{%
\stepcounter{tutetotcount}%
\stepcounter{tutecount}%
\ifnum\value{tutecount}=27%
\stepcounter{tutetutecount}%
\addtocounter{tutecount}{-26}%
\fi%
\xdef#1{{\ifnum\value{tutetutecount}>0\alph{tutetutecount}\fi\alph{tutecount}}}}
\def\tute#1{$^{#1}$\stepcounter{namecount}}
\newcounter{notecount}
\newcommand{\note}{{\stepcounter{notecount}\thenotecount}}
\tutenum\aachenI           
\tutenum\aachenIII            
\tutenum\lapp              
\tutenum\jhu
\tutenum\lsu
\tutenum\cssa
\tutenum\calt
\tutenum\iee
\tutenum\ihep           
\tutenum\bologna           
\tutenum\bucharest         
\tutenum\mit               
\tutenum\ncu               
\tutenum\coimbra
\tutenum\maryland
\tutenum\florence          
\tutenum\mpi
\tutenum\geneva            
\tutenum\grenoble
\tutenum\hefei
\tutenum\hut
\tutenum\ist
\tutenum\lip
\tutenum\csist
\tutenum\madrid            
\tutenum\milan             
\tutenum\kurch
\tutenum\moscow            
\tutenum\perugia           
\tutenum\as
\tutenum\korea
\tutenum\turku
\tutenum\eth               
{
\parskip=0pt
\section*{The AMS Collaboration}
\tolerance=10000
\hbadness=5000
\raggedright
\def\r{\rlap,}
\noindent
J.Alcaraz\r\tute\madrid\
D.Alvisi\r\tute\bologna\
B.Alpat\r\tute\perugia\
G.Ambrosi\r\tute\geneva\
H.Anderhub\r\tute\eth\
L.Ao\r\tute\calt\
A.Arefiev\r\tute\moscow\
P.Azzarello\r\tute\geneva\
E.Babucci\r\tute\perugia\
L.Baldini\r\tute{\bologna,\mit}\
% ...
\typeout{--------------------------------------------------------------}
\typeout{
Imagine that:  <\thenamecount> authors from <\thetutetotcount> institutes.}
\typeout{--------------------------------------------------------------}
\vspace*{-.5\baselineskip}
\rule[.5\baselineskip]{\textwidth}{0.5pt}
\begin{list}{A}{\itemsep=0pt plus 0pt minus 0pt\parsep=0pt plus 0pt minus 0pt
                \topsep=0pt plus 0pt minus 0pt}
\small
\item[$^\aachenI$]
 I. Physikalisches Institut, RWTH, D-52056 Aachen, Germany$^\note$
\item[$^\aachenIII$]
 III. Physikalisches Institut, RWTH, D-52056 Aachen, Germany$^6$
\item[$^\lapp$] Laboratoire d'Annecy-le-Vieux de Physique des Particules, 
     LAPP, F-74941 Annecy-le-Vieux CEDEX, France
\item[$^\lsu$] Louisiana State University, Baton Rouge, LA 70803, USA
\item[$^\jhu$] Johns Hopkins University, Baltimore, MD 21218, USA
\item[$^\cssa$] Center of Space Science and Application, 
  Chinese Academy of Sciences,
  100080 Beijing, China
\item[$^\calt$] Chinese Academy of Launching Vehicle Technology, CALT,
  100076 Beijing, China
\item[$^\iee$] Institute of Electrical Engineering, IEE, 
  Chinese Academy of Sciences, 100080 Beijing, China
\item[$^\ihep$] Institute of High Energy Physics, IHEP, 
  Chinese Academy of Sciences, 
  100039 Beijing, China$^\note$      
\item[$^\bologna$] University of Bologna and  INFN-Sezione di Bologna, 
     I-40126 Bologna, Italy 
\item[$^\bucharest$] Institute of Microtechnology, 
                    Politechnica University of Bucharest 
                    and University of Bucharest,
                    R-76900 Bucharest, Romania
\item[$^\mit$] Massachusetts Institute of Technology, Cambridge, MA 02139, USA
\item[$^\ncu$] National Central University, Chung-Li, Taiwan 32054
\item[$^\coimbra$]  Laboratorio de Instrumentacao e Fisica Experimental de 
            Particulas, LIP, P-3000 Coimbra, Portugal
\item[$^\maryland$] University of Maryland, College Park, MD 20742, USA
\item[$^\florence$] INFN Sezione di Firenze, 
     I-50125 Florence, Italy
\item[$^\mpi$] Max--Plank Institut fur Extraterrestrische Physik,
            D-85740 Garching, Germany
\item[$^\geneva$] University of Geneva, CH-1211 Geneva 4, Switzerland
\item[$^\grenoble$] Institut des Sciences Nucleaires,
     F-38026 Grenoble, France
\item[$^\hefei$] Chinese University of Science and Technology, USTC,
      Hefei, Anhui 230 029, China$^{7}$
\item[$^\hut$] Helsinki University of Technology,
    FIN-02540 Kylmala, Finland
\item[$^\ist$] Instituto Superior T\'ecnico, IST,  P-1096 Lisboa, Portugal
\item[$^\lip$] Laboratorio de Instrumentacao e Fisica Experimental de 
            Particulas, LIP, P-1000 Lisboa, Portugal
\item[$^\csist$] Chung--Shan Institute of Science and Technology,
     Lung-Tan, Tao Yuan 325, Taiwan 11529
\item[$^\madrid$] Centro de Investigaciones Energ{\'e}ticas, 
     Medioambientales y Tecnolog{\'\i}cas, CIEMAT, E-28040 Madrid,
     Spain$^\note$ 
\item[$^\milan$] INFN-Sezione di Milano, I-20133 Milan, Italy
\item[$^\kurch$] Kurchatov Institute, Moscow, 123182 Russia
\item[$^\moscow$] Institute of Theoretical and Experimental Physics, ITEP, 
     Moscow, 117259 Russia  
\item[$^\perugia$] INFN-Sezione di Perugia and Universit\'a Degli 
     Studi di Perugia, I-06100 Perugia, Italy$^\note$  
\item[$^\as$] Academia Sinica,
    Taipei, Taiwan
\item[$^\korea$] Kyungpook National University, 
     702-701 Taegu, Korea
\item[$^\turku$] University of Turku,
    FIN-20014 Turku, Finland
\item[$^\eth$] Eidgen\"ossische Technische Hochschule, ETH Z\"urich,
     CH-8093 Z\"urich, Switzerland
\setcounter{notecount}{0}
%
%
\item[$^\note$] Permanent address:  HEPPG, Univ.~of Bucharest, Romania.
%
\item[$^\note$] Permanent address: Nuclear Physics Institute, 
               St. Petersburg, Russia.
%
\item[$^\note$] Now at European Laboratory for Particle Physics, CERN, 
     CH-1211 Geneva 23, Switzerland.
%
\item[$^\note$] Now at National Institute for High Energy Physics, NIKHEF, 
           NL-1009 DB Amsterdam, The Netherlands.
%
\item[$^\note$] Supported by ETH Z\"urich.
%
%
\item[$^\note$]  Supported by the 
Deutsches Zentrum f\"ur Luft-- und Raumfahrt, DLR.
%
\item[$^\note$] Supported by the National Natural Science Foundation of China.
%
\item[$^\note$] Supported also by the Comisi\'on Interministerial de Ciencia y 
           Tecnolog{\'\i}a.
%
\item[$^\note$] Also supported by the Italian Space Agency.
\end{list}

}
%
%
%
\newcommand{\X}{$\raisebox{1pt}{\scriptsize$\,\times$}$}
\newcommand{\PM}{\!\pm\!}

\newpage
\begin{table}[hp]
\begin{center}
\renewcommand{\arraystretch}{1.20}
\small
\begin{tabular}{|r@{\,--\,}r|c@{\,\vline\,\,}c@{\,\vline\,\,}c@{\,\vline\,\,}%
c@{\,\vline\,\,}c@{\,\vline}}
\hline
\multicolumn{7}{|c@{\,\vline}}{\Large $\strut$ Downward Proton Flux 
(m$^{2}$\,sec\,sr\,\MeV)$^{-1}$}\\\hline
\multicolumn{2}{|c|}{$E_{kin}$}&%
\multicolumn{5}{c@{\,\vline}}{Geomagnetic Latitude Range}\\
\cline{3-7}
\multicolumn{2}{|c|}{$(\GeV)$}&%
$\TM<0.2$&$0.2\le\TM<0.3$&$0.3\le\TM<0.4$&$0.4\le\TM<0.5$&$0.5\le\TM<0.6$\\
\hline
  0.07&  0.10&$(16.7\PM4.4)\X10^{-2}$&$(14.2\PM4.0)\X10^{-2}$&$(11.2\PM3.1)\X10^{-2}$&$(13.6\PM3.8)\X10^{-2}$&$(13.4\PM3.6)\X10^{-2}$\\
  0.10&  0.15&$(12.1\PM1.4)\X10^{-2}$&$(\pho8.2\PM1.0)\X10^{-2}$&$(\pho7.6\PM1.0)\X10^{-2}$&$(\pho7.6\PM1.0)\X10^{-2}$&$(\pho7.7\PM1.0)\X10^{-2}$\\
  0.15&  0.22&$(97.9\PM4.6)\X10^{-3}$&$(51.2\PM3.2)\X10^{-3}$&$(41.9\PM2.6)\X10^{-3}$&$(44.6\PM3.0)\X10^{-3}$&$(48.4\PM3.3)\X10^{-3}$\\
  0.22&  0.31&$(86.2\PM2.8)\X10^{-3}$&$(45.6\PM1.8)\X10^{-3}$&$(37.9\PM1.7)\X10^{-3}$&$(34.4\PM1.5)\X10^{-3}$&$(32.7\PM1.6)\X10^{-3}$\\\hline
  0.31&  0.44&$(70.1\PM3.2)\X10^{-3}$&$(34.6\PM1.5)\X10^{-3}$&$(24.4\PM1.1)\X10^{-3}$&$(21.1\PM1.2)\X10^{-3}$&$(20.2\PM1.2)\X10^{-3}$\\
  0.44&  0.62&$(50.4\PM2.7)\X10^{-3}$&$(21.2\PM1.2)\X10^{-3}$&$(155.\PM9.3)\X10^{-4}$&$(121.\PM9.3)\X10^{-4}$&$(113.\PM9.0)\X10^{-4}$\\
  0.62&  0.85&$(32.8\PM1.9)\X10^{-3}$&$(116.\PM6.8)\X10^{-4}$&$(84.9\PM6.5)\X10^{-4}$&$(61.5\PM5.6)\X10^{-4}$&$(50.0\PM6.4)\X10^{-4}$\\
  0.85&  1.15&$(20.6\PM1.2)\X10^{-3}$&$(57.2\PM4.7)\X10^{-4}$&$(40.0\PM3.8)\X10^{-4}$&$(26.9\PM3.4)\X10^{-4}$&$(24.2\PM4.2)\X10^{-4}$\\\hline
  1.15&  1.54&$(116.\PM6.9)\X10^{-4}$&$(28.6\PM3.3)\X10^{-4}$&$(17.7\PM2.5)\X10^{-4}$&$(12.7\PM2.9)\X10^{-4}$&$(\pho8.5\PM1.4)\X10^{-4}$\\
  1.54&  2.02&$(66.9\PM4.2)\X10^{-4}$&$(12.2\PM2.1)\X10^{-4}$&$(\pho8.5\PM2.6)\X10^{-4}$&$(\pho6.9\PM1.4)\X10^{-4}$&$(\pho5.7\PM1.0)\X10^{-4}$\\
  2.02&  2.62&$(28.6\PM1.9)\X10^{-4}$&$(\pho8.2\PM1.8)\X10^{-4}$&$(\pho5.0\PM1.3)\X10^{-4}$&$(37.3\PM3.3)\X10^{-5}$&$(34.2\PM1.5)\X10^{-5}$\\
  2.62&  3.38&$(110.\PM9.6)\X10^{-5}$&$(\pho3.6\PM1.1)\X10^{-4}$&$(30.0\PM8.6)\X10^{-5}$&$(204.\PM7.4)\X10^{-6}$&$(29.0\PM1.4)\X10^{-5}$\\\hline
  3.38&  4.31&$(44.3\PM7.9)\X10^{-5}$&$(20.3\PM6.0)\X10^{-5}$&$(23.2\PM3.6)\X10^{-5}$&$(25.0\PM1.3)\X10^{-5}$&$(10.7\PM1.1)\X10^{-4}$\\
  4.31&  5.45&$(15.7\PM3.1)\X10^{-5}$&$(13.4\PM4.8)\X10^{-5}$&$(17.6\PM3.2)\X10^{-5}$&$(58.5\PM5.9)\X10^{-5}$&$(62.9\PM6.4)\X10^{-4}$\\
  5.45&  6.86&$(\pho6.1\PM2.2)\X10^{-5}$&$(105.\PM8.7)\X10^{-6}$&$(31.9\PM2.3)\X10^{-5}$&$(32.1\PM3.0)\X10^{-4}$&$(18.4\PM1.4)\X10^{-3}$\\
  6.86&  8.60&$(23.7\PM2.1)\X10^{-5}$&$(53.8\PM2.7)\X10^{-5}$&$(19.5\PM1.5)\X10^{-4}$&$(96.2\PM6.4)\X10^{-4}$&$(23.3\PM1.2)\X10^{-3}$\\\hline
  8.60& 10.73&$(138.\PM6.8)\X10^{-5}$&$(28.6\PM1.7)\X10^{-4}$&$(58.5\PM3.3)\X10^{-4}$&$(128.\PM5.4)\X10^{-4}$&$(193.\PM5.1)\X10^{-4}$\\
 10.73& 13.34&$(49.5\PM1.8)\X10^{-4}$&$(60.9\PM2.4)\X10^{-4}$&$(85.7\PM3.1)\X10^{-4}$&$(115.\PM2.8)\X10^{-4}$&$(128.\PM3.7)\X10^{-4}$\\
 13.34& 16.55&$(65.7\PM2.1)\X10^{-4}$&$(63.4\PM1.8)\X10^{-4}$&$(72.1\PM2.1)\X10^{-4}$&$(75.6\PM2.5)\X10^{-4}$&$(75.6\PM2.7)\X10^{-4}$\\
 16.55& 20.48&$(45.7\PM1.7)\X10^{-4}$&$(45.5\PM1.7)\X10^{-4}$&$(44.4\PM1.5)\X10^{-4}$&$(45.2\PM1.8)\X10^{-4}$&$(43.3\PM1.2)\X10^{-4}$\\\hline
 20.48& 25.29&$(27.7\PM1.0)\X10^{-4}$&$(25.5\PM1.0)\X10^{-4}$&$(255.\PM9.8)\X10^{-5}$&$(248.\PM9.6)\X10^{-5}$&$(24.0\PM1.0)\X10^{-4}$\\
 25.29& 31.20&$(155.\PM5.9)\X10^{-5}$&$(147.\PM7.1)\X10^{-5}$&$(144.\PM6.8)\X10^{-5}$&$(142.\PM6.7)\X10^{-5}$&$(138.\PM5.6)\X10^{-5}$\\
 31.20& 38.43&$(90.5\PM4.1)\X10^{-5}$&$(79.2\PM4.7)\X10^{-5}$&$(80.5\PM4.5)\X10^{-5}$&$(80.0\PM4.3)\X10^{-5}$&$(77.1\PM4.3)\X10^{-5}$\\
 38.43& 47.30&$(51.4\PM2.2)\X10^{-5}$&$(48.9\PM3.0)\X10^{-5}$&$(48.2\PM2.5)\X10^{-5}$&$(48.2\PM3.0)\X10^{-5}$&$(47.1\PM2.7)\X10^{-5}$\\\hline
 47.30& 58.16&$(30.0\PM1.7)\X10^{-5}$&$(28.6\PM2.0)\X10^{-5}$&$(28.7\PM1.8)\X10^{-5}$&$(28.4\PM1.8)\X10^{-5}$&$(27.7\PM1.8)\X10^{-5}$\\
 58.16& 71.48&$(164.\PM8.8)\X10^{-6}$&$(15.4\PM1.2)\X10^{-5}$&$(15.6\PM1.2)\X10^{-5}$&$(154.\PM8.8)\X10^{-6}$&$(149.\PM9.9)\X10^{-6}$\\
 71.48& 87.79&$(86.1\PM3.9)\X10^{-6}$&$(79.6\PM4.7)\X10^{-6}$&$(81.5\PM6.4)\X10^{-6}$&$(80.2\PM5.9)\X10^{-6}$&$(76.7\PM5.1)\X10^{-6}$\\
 87.79&107.78&$(49.4\PM2.9)\X10^{-6}$&$(45.0\PM4.6)\X10^{-6}$&$(46.6\PM4.8)\X10^{-6}$&$(45.8\PM2.8)\X10^{-6}$&$(43.4\PM2.6)\X10^{-6}$\\\hline
107.78&132.27&$(28.6\PM3.1)\X10^{-6}$&$(25.7\PM6.1)\X10^{-6}$&$(26.9\PM7.3)\X10^{-6}$&$(26.4\PM6.2)\X10^{-6}$&$(24.8\PM4.6)\X10^{-6}$\\
132.27&162.29&$(16.2\PM1.8)\X10^{-6}$&$(14.3\PM7.0)\X10^{-6}$&$(15.2\PM5.2)\X10^{-6}$&$(14.9\PM7.9)\X10^{-6}$&$(13.8\PM6.3)\X10^{-6}$\\
162.29&199.06&$(97.2\PM5.1)\X10^{-7}$&$(84.8\PM6.7)\X10^{-7}$&$(\pho9.1\PM2.3)\X10^{-6}$&$(\pho8.9\PM1.8)\X10^{-6}$&$(82.1\PM6.2)\X10^{-7}$\\
\hline
\end{tabular}
\end{center}  
\caption{Differential downward proton flux spectra for lower latitudes.
\label{fluxdownlo}}
\end{table}

\newpage
\begin{table}[hp]
\begin{center}
\renewcommand{\arraystretch}{1.20}
\small
\begin{tabular}{|r@{\,--\,}r|c@{\,\vline\,\,}c@{\,\vline\,\,}c@{\,\vline\,\,}%
c@{\,\vline\,\,}c@{\,\vline}}
\hline
\multicolumn{7}{|c@{\,\vline}}{\Large $\strut$ Downward Proton Flux 
(m$^{2}$\,sec\,sr\,\MeV)$^{-1}$}\\\hline
\multicolumn{2}{|c|}{$E_{kin}$}&%
\multicolumn{5}{c@{\,\vline}}{Geomagnetic Latitude Range}\\
\cline{3-7}
\multicolumn{2}{|c|}{$(\GeV)$}&%
$0.6\le\TM<0.7$&$0.7\le\TM<0.8$&$0.8\le\TM<0.9$&$0.9\le\TM<1.0$&$1.0\le\TM$\\
\hline
  0.07&  0.10&$(12.2\PM3.5)\X10^{-2}$&$(18.5\PM5.9)\X10^{-2}$&$(25.1\PM8.9)\X10^{-2}$&$(\pho4.3\PM1.3)\X10^{-1}$&$(\pho9.2\PM2.6)\X10^{-1}$\\
  0.10&  0.15&$(\pho9.7\PM1.3)\X10^{-2}$&$(11.8\PM1.6)\X10^{-2}$&$(19.1\PM2.6)\X10^{-2}$&$(41.8\PM5.6)\X10^{-2}$&$(\pho9.8\PM1.2)\X10^{-1}$\\
  0.15&  0.22&$(66.0\PM3.7)\X10^{-3}$&$(97.3\PM5.9)\X10^{-3}$&$(144.\PM8.9)\X10^{-3}$&$(33.6\PM3.3)\X10^{-2}$&$(109.\PM6.7)\X10^{-2}$\\
  0.22&  0.31&$(44.4\PM1.6)\X10^{-3}$&$(44.2\PM2.0)\X10^{-3}$&$(92.4\PM6.9)\X10^{-3}$&$(22.6\PM3.9)\X10^{-2}$&$(126.\PM5.3)\X10^{-2}$\\\hline
  0.31&  0.44&$(24.1\PM1.7)\X10^{-3}$&$(23.8\PM1.3)\X10^{-3}$&$(58.3\PM4.8)\X10^{-3}$&$(29.3\PM7.1)\X10^{-2}$&$(139.\PM4.1)\X10^{-2}$\\
  0.44&  0.62&$(108.\PM8.8)\X10^{-4}$&$(14.4\PM1.0)\X10^{-3}$&$(36.6\PM3.5)\X10^{-3}$&$(\pho4.7\PM1.1)\X10^{-1}$&$(132.\PM4.8)\X10^{-2}$\\
  0.62&  0.85&$(47.8\PM6.7)\X10^{-4}$&$(77.2\PM6.9)\X10^{-4}$&$(22.0\PM2.5)\X10^{-3}$&$(\pho7.5\PM1.3)\X10^{-1}$&$(114.\PM4.2)\X10^{-2}$\\
  0.85&  1.15&$(23.1\PM4.9)\X10^{-4}$&$(60.9\PM6.5)\X10^{-4}$&$(34.9\PM5.8)\X10^{-3}$&$(85.3\PM7.5)\X10^{-2}$&$(92.8\PM3.2)\X10^{-2}$\\\hline
  1.15&  1.54&$(13.1\PM2.2)\X10^{-4}$&$(23.7\PM2.9)\X10^{-4}$&$(15.4\PM2.4)\X10^{-2}$&$(71.7\PM4.5)\X10^{-2}$&$(72.4\PM2.4)\X10^{-2}$\\
  1.54&  2.02&$(\pho7.7\PM1.2)\X10^{-4}$&$(44.8\PM6.7)\X10^{-4}$&$(28.1\PM3.3)\X10^{-2}$&$(52.4\PM4.5)\X10^{-2}$&$(51.1\PM1.4)\X10^{-2}$\\
  2.02&  2.62&$(77.7\PM8.3)\X10^{-5}$&$(43.1\PM5.8)\X10^{-3}$&$(30.9\PM1.8)\X10^{-2}$&$(36.2\PM2.9)\X10^{-2}$&$(37.0\PM1.1)\X10^{-2}$\\
  2.62&  3.38&$(49.1\PM5.9)\X10^{-4}$&$(11.4\PM1.1)\X10^{-2}$&$(22.6\PM1.4)\X10^{-2}$&$(24.8\PM2.1)\X10^{-2}$&$(241.\PM6.4)\X10^{-3}$\\\hline
  3.38&  4.31&$(27.9\PM2.9)\X10^{-3}$&$(124.\PM4.6)\X10^{-3}$&$(15.4\PM1.1)\X10^{-2}$&$(16.2\PM1.1)\X10^{-2}$&$(163.\PM3.1)\X10^{-3}$\\
  4.31&  5.45&$(56.4\PM4.0)\X10^{-3}$&$(88.4\PM4.3)\X10^{-3}$&$(95.3\PM5.9)\X10^{-3}$&$(103.\PM7.7)\X10^{-3}$&$(102.\PM2.9)\X10^{-3}$\\
  5.45&  6.86&$(52.6\PM1.7)\X10^{-3}$&$(55.6\PM3.2)\X10^{-3}$&$(59.3\PM3.5)\X10^{-3}$&$(63.8\PM5.0)\X10^{-3}$&$(61.4\PM1.3)\X10^{-3}$\\
  6.86&  8.60&$(35.6\PM1.2)\X10^{-3}$&$(34.0\PM1.8)\X10^{-3}$&$(36.3\PM2.6)\X10^{-3}$&$(39.0\PM2.8)\X10^{-3}$&$(390.\PM8.2)\X10^{-4}$\\\hline
  8.60& 10.73&$(212.\PM9.0)\X10^{-4}$&$(20.2\PM1.1)\X10^{-3}$&$(21.8\PM1.6)\X10^{-3}$&$(22.5\PM1.6)\X10^{-3}$&$(223.\PM6.5)\X10^{-4}$\\
 10.73& 13.34&$(129.\PM5.3)\X10^{-4}$&$(121.\PM6.4)\X10^{-4}$&$(128.\PM8.0)\X10^{-4}$&$(14.1\PM1.3)\X10^{-3}$&$(136.\PM4.5)\X10^{-4}$\\
 13.34& 16.55&$(75.8\PM3.3)\X10^{-4}$&$(69.0\PM3.8)\X10^{-4}$&$(75.2\PM4.3)\X10^{-4}$&$(78.0\PM5.7)\X10^{-4}$&$(76.2\PM2.7)\X10^{-4}$\\
 16.55& 20.48&$(41.7\PM1.5)\X10^{-4}$&$(40.5\PM2.1)\X10^{-4}$&$(40.2\PM3.0)\X10^{-4}$&$(39.3\PM3.3)\X10^{-4}$&$(39.6\PM1.3)\X10^{-4}$\\\hline
 20.48& 25.29&$(24.9\PM1.1)\X10^{-4}$&$(22.7\PM1.3)\X10^{-4}$&$(237.\PM8.0)\X10^{-5}$&$(23.8\PM2.0)\X10^{-4}$&$(22.0\PM1.3)\X10^{-4}$\\
 25.29& 31.20&$(134.\PM5.6)\X10^{-5}$&$(132.\PM8.7)\X10^{-5}$&$(127.\PM6.4)\X10^{-5}$&$(12.3\PM1.4)\X10^{-4}$&$(118.\PM7.9)\X10^{-5}$\\
 31.20& 38.43&$(75.1\PM4.0)\X10^{-5}$&$(69.2\PM4.5)\X10^{-5}$&$(61.5\PM5.7)\X10^{-5}$&$(78.0\PM8.8)\X10^{-5}$&$(76.7\PM6.5)\X10^{-5}$\\
 38.43& 47.30&$(46.0\PM2.7)\X10^{-5}$&$(44.7\PM2.8)\X10^{-5}$&$(44.0\PM3.5)\X10^{-5}$&$(44.1\PM4.6)\X10^{-5}$&$(47.7\PM3.7)\X10^{-5}$\\\hline
 47.30& 58.16&$(27.0\PM1.8)\X10^{-5}$&$(26.3\PM1.9)\X10^{-5}$&$(25.7\PM2.8)\X10^{-5}$&$(27.0\PM2.6)\X10^{-5}$&$(28.5\PM2.6)\X10^{-5}$\\
 58.16& 71.48&$(14.6\PM1.2)\X10^{-5}$&$(142.\PM9.9)\X10^{-6}$&$(13.9\PM1.3)\X10^{-5}$&$(14.3\PM1.5)\X10^{-5}$&$(154.\PM9.8)\X10^{-6}$\\
 71.48& 87.79&$(76.0\PM4.6)\X10^{-6}$&$(72.9\PM4.5)\X10^{-6}$&$(71.7\PM6.4)\X10^{-6}$&$(72.5\PM6.5)\X10^{-6}$&$(79.3\PM8.7)\X10^{-6}$\\
 87.79&107.78&$(43.5\PM5.8)\X10^{-6}$&$(41.5\PM3.0)\X10^{-6}$&$(41.1\PM4.1)\X10^{-6}$&$(40.3\PM6.3)\X10^{-6}$&$(44.8\PM7.9)\X10^{-6}$\\\hline
107.78&132.27&$(25.2\PM4.5)\X10^{-6}$&$(23.9\PM4.4)\X10^{-6}$&$(23.9\PM4.4)\X10^{-6}$&$(\pho2.3\PM1.2)\X10^{-5}$&$(\pho2.6\PM1.2)\X10^{-5}$\\
132.27&162.29&$(14.3\PM3.9)\X10^{-6}$&$(13.4\PM4.7)\X10^{-6}$&$(13.6\PM6.5)\X10^{-6}$&$(12.3\PM8.9)\X10^{-6}$&$(\pho1.4\PM1.4)\X10^{-5}$\\
162.29&199.06&$(\pho8.6\PM1.5)\X10^{-6}$&$(80.6\PM4.3)\X10^{-7}$&$(\pho8.2\PM1.3)\X10^{-6}$&$(\pho7.2\PM3.7)\X10^{-6}$&$(\pho8.5\PM2.4)\X10^{-6}$\\
\hline
\end{tabular}
\end{center}  
\caption{Differential downward proton flux spectra for higher latitudes.
\label{fluxdownhi}}
\end{table}

\newpage
\begin{table}[hp]
\begin{center}
\renewcommand{\arraystretch}{1.20}
\small
\begin{tabular}{|r@{\,--\,}r|c@{\,\vline\,\,}c@{\,\vline\,\,}c@{\,\vline\,\,}%
c@{\,\vline\,\,}c@{\,\vline}}
\hline
\multicolumn{7}{|c@{\,\vline}}{\Large $\strut$ Upward Proton Flux 
(m$^{2}$\,sec\,sr\,\MeV)$^{-1}$}\\\hline
\multicolumn{2}{|c|}{$E_{kin}$}&%
\multicolumn{5}{c@{\,\vline}}{Geomagnetic Latitude Range}\\
\cline{3-7}
\multicolumn{2}{|c|}{$(\GeV)$}&%
$\TM<0.2$&$0.2\le\TM<0.3$&$0.3\le\TM<0.4$&$0.4\le\TM<0.5$&$0.5\le\TM<0.6$\\
\hline
  0.07&  0.10&$(16.4\PM4.4)\X10^{-2}$&$(13.1\PM3.9)\X10^{-2}$&$(12.6\PM3.5)\X10^{-2}$&$(14.7\PM4.1)\X10^{-2}$&$(15.8\PM4.7)\X10^{-2}$\\
  0.10&  0.15&$(10.9\PM1.4)\X10^{-2}$&$(\pho7.5\PM1.0)\X10^{-2}$&$(66.0\PM9.2)\X10^{-3}$&$(\pho7.7\PM1.1)\X10^{-2}$&$(\pho8.7\PM1.2)\X10^{-2}$\\
  0.15&  0.22&$(85.3\PM4.9)\X10^{-3}$&$(48.1\PM3.5)\X10^{-3}$&$(42.7\PM2.8)\X10^{-3}$&$(42.2\PM2.8)\X10^{-3}$&$(46.3\PM2.8)\X10^{-3}$\\
  0.22&  0.31&$(84.8\PM3.8)\X10^{-3}$&$(44.5\PM2.1)\X10^{-3}$&$(39.3\PM1.9)\X10^{-3}$&$(35.5\PM1.8)\X10^{-3}$&$(34.6\PM1.5)\X10^{-3}$\\\hline
  0.31&  0.44&$(66.8\PM3.4)\X10^{-3}$&$(33.6\PM1.7)\X10^{-3}$&$(25.4\PM1.1)\X10^{-3}$&$(21.4\PM1.1)\X10^{-3}$&$(21.0\PM1.1)\X10^{-3}$\\
  0.44&  0.62&$(48.4\PM2.7)\X10^{-3}$&$(20.3\PM1.2)\X10^{-3}$&$(136.\PM8.3)\X10^{-4}$&$(124.\PM9.2)\X10^{-4}$&$(97.6\PM8.1)\X10^{-4}$\\
  0.62&  0.85&$(32.7\PM2.0)\X10^{-3}$&$(120.\PM8.6)\X10^{-4}$&$(76.4\PM5.6)\X10^{-4}$&$(61.9\PM6.1)\X10^{-4}$&$(34.8\PM4.3)\X10^{-4}$\\
  0.85&  1.15&$(20.2\PM1.1)\X10^{-3}$&$(53.9\PM4.6)\X10^{-4}$&$(42.0\PM4.5)\X10^{-4}$&$(31.9\PM4.6)\X10^{-4}$&$(17.9\PM3.3)\X10^{-4}$\\\hline
  1.15&  1.54&$(124.\PM7.1)\X10^{-4}$&$(34.8\PM4.4)\X10^{-4}$&$(14.7\PM1.8)\X10^{-4}$&$(14.0\PM2.3)\X10^{-4}$&$(\pho8.6\PM2.1)\X10^{-4}$\\
  1.54&  2.02&$(62.0\PM4.2)\X10^{-4}$&$(16.4\PM2.3)\X10^{-4}$&$(12.5\PM2.3)\X10^{-4}$&$(\pho8.8\PM1.8)\X10^{-4}$&$(\pho5.2\PM1.2)\X10^{-4}$\\
  2.02&  2.62&$(25.9\PM1.8)\X10^{-4}$&$(\pho7.9\PM1.3)\X10^{-4}$&$(\pho5.6\PM1.1)\X10^{-4}$&$(\pho4.6\PM1.2)\X10^{-4}$&$(\pho3.4\PM1.1)\X10^{-4}$\\
  2.62&  3.38&$(10.7\PM1.5)\X10^{-4}$&$(\pho4.2\PM1.2)\X10^{-4}$&$(29.9\PM8.7)\X10^{-5}$&$(38.3\PM10.)\X10^{-5}$&$(25.9\PM9.6)\X10^{-5}$\\\hline
  3.38&  4.31&$(29.7\PM5.7)\X10^{-5}$&$(15.6\PM8.3)\X10^{-5}$&$(11.9\PM4.9)\X10^{-5}$&$(13.4\PM5.7)\X10^{-5}$&$(\pho9.4\PM3.7)\X10^{-5}$\\
  4.31&  5.45&$(11.2\PM4.6)\X10^{-5}$&$(\pho6.4\PM4.2)\X10^{-5}$&$(\pho7.2\PM3.8)\X10^{-5}$&$(\pho6.4\PM3.3)\X10^{-5}$&                       \\
  5.45&  6.86&$(\pho3.7\PM2.4)\X10^{-5}$&                       &                          &                          &                       \\
\hline
\hline
\multicolumn{2}{|c|}{$E_{kin}$}&%
\multicolumn{5}{c@{\,\vline}}{Geomagnetic Latitude Range}\\
\cline{3-7}
\multicolumn{2}{|c|}{$(\GeV)$}&%
$0.6\le\TM<0.7$&$0.7\le\TM<0.8$&$0.8\le\TM<0.9$&$0.9\le\TM<1.0$&\\
\hline
  0.07&  0.10&$(23.1\PM6.8)\X10^{-2}$&$(32.9\PM9.5)\X10^{-2}$&$(\pho3.8\PM1.1)\X10^{-1}$&$(\pho5.1\PM1.5)\X10^{-1}$&\\
  0.10&  0.15&$(10.5\PM1.5)\X10^{-2}$&$(15.4\PM2.3)\X10^{-2}$&$(18.0\PM2.4)\X10^{-2}$&$(25.5\PM4.1)\X10^{-2}$&\\
  0.15&  0.22&$(58.1\PM3.8)\X10^{-3}$&$(72.5\PM5.4)\X10^{-3}$&$(91.9\PM6.2)\X10^{-3}$&$(99.8\PM8.4)\X10^{-3}$&\\
  0.22&  0.31&$(43.0\PM2.1)\X10^{-3}$&$(44.8\PM3.4)\X10^{-3}$&$(57.4\PM3.3)\X10^{-3}$&$(54.0\PM4.9)\X10^{-3}$&\\\hline
  0.31&  0.44&$(20.7\PM1.1)\X10^{-3}$&$(21.7\PM1.9)\X10^{-3}$&$(25.7\PM2.6)\X10^{-3}$&$(22.5\PM2.9)\X10^{-3}$&\\
  0.44&  0.62&$(83.4\PM8.0)\X10^{-4}$&$(78.6\PM9.3)\X10^{-4}$&$(\pho8.8\PM1.2)\X10^{-3}$&$(\pho8.8\PM1.7)\X10^{-3}$&\\
  0.62&  0.85&$(27.3\PM4.0)\X10^{-4}$&$(18.4\PM3.2)\X10^{-4}$&$(17.9\PM4.8)\X10^{-4}$&$(23.4\PM8.0)\X10^{-4}$&\\
  0.85&  1.15&$(\pho7.2\PM2.3)\X10^{-4}$&$(\pho4.9\PM1.9)\X10^{-4}$&$(\pho7.4\PM4.2)\X10^{-4}$&$(12.6\PM5.1)\X10^{-4}$&\\\hline
  1.15&  1.54&$(\pho4.0\PM1.3)\X10^{-4}$&$(\pho3.2\PM2.3)\X10^{-4}$&$(\pho2.5\PM1.5)\X10^{-4}$&$(\pho9.1\PM4.0)\X10^{-4}$&\\
  1.54&  2.02&$(\pho3.0\PM1.4)\X10^{-4}$&$(11.6\PM7.2)\X10^{-5}$&$(\pho1.3\PM1.2)\X10^{-4}$&$(16.8\PM9.3)\X10^{-5}$&\\
  2.02&  2.62&$(\pho1.7\PM1.2)\X10^{-4}$&$(\pho7.7\PM7.4)\X10^{-5}$&                 & &\\
  2.62&  3.38&$(\pho6.3\PM4.1)\X10^{-5}$&$(\pho4.8\PM3.8)\X10^{-5}$&                 & &\\\hline
  3.38&  4.31&$(\pho2.0\PM1.1)\X10^{-5}$&                    &                       &                       &\\
\hline
\end{tabular}
\end{center}
\caption {Differential upward proton flux spectra.
\label{fluxup}}
\end{table}
%

\clearpage
\newpage

\clearpage
\newpage
 
\clearpage
\newpage

\clearpage
\newpage 

\clearpage
\newpage 

\clearpage
\newpage

\clearpage
\newpage

\end{document}